\def\PR{{Phys.~Rev.~}}
\def\PRL{{ Phys.~Rev.~Lett.~}}
\def\PRB{{ Phys.~Rev.~B~}}
\def\PRA{{ Phys.~Rev.~A~}}

\def\etal{{\it et al.}}
\documentclass[aps,prb,showpacs,superscriptaddress,a4paper,twocolumn]{revtex4-1}

\usepackage{graphicx}

\begin {document}

\author{Prashant Singh}\email{prashant@ameslab.gov}
\affiliation{Ames Laboratory, U.S. Department of Energy, Iowa State University, Ames, Iowa 50011-3020, USA}
\author{Manoj K Harbola}\email{mkh@iitk.ac.in}
\affiliation{Department of Physics, Indian Institute of Technology, Kanpur, 208016, India}
\author{Duane D. Johnson}\email{ddj@iastate.edu; ddj@ameslab.gov}
\affiliation{Ames Laboratory, U.S. Department of Energy, Iowa State University, Ames, Iowa 50011-3020, USA}
\affiliation{Materials Science \& Engineering, Iowa State University, Ames, Iowa 50011-2300, USA} 

\title{Better band gaps for wide-gap semiconductors from a locally corrected exchange-correlation potential that nearly eliminates self-interaction errors}

\begin{abstract}
This work constitutes a comprehensive and improved account of electronic-structure and mechanical properties of silicon-nitride (Si$_{3}$N$_{4}$) polymorphs via van Leeuwen and Baerends (LB) exchange-corrected local density approximation (LDA) that enforces the exact exchange potential asymptotic behavior. The calculated lattice constant, bulk modulus, and electronic band structure of Si$_{3}$N$_{4}$ polymorphs are in good agreement with experimental results. We also show that, for a single electron in a hydrogen atom,  spherical well, or harmonic oscillator, the LB-corrected LDA reduces the (self-interaction) error to exact total energy to $\sim$10\%,  a factor of 3 to 4 lower than standard LDA, due to a dramatically improved representation of the exchange-potential. 
\end{abstract}
\date{\today}

\maketitle

\section{Introduction}
%The research on silicon-nitride has largely been driven by its use in microelectronics technology to utilize it as an effective insulating material and also as diffusion mask for impurities.\cite{1991SiCBook}  
The wide-band gap silicon nitride (Si$_{3}$N$_{4}$) is an important engineering ceramic material owing to its superior physical and chemical properties, high chemical and thermal stability, and good wear resistance. It has been widely applied in cutting tools, engine components, ball bearings and gas turbines.\cite{1,2,3,1999NatureX3N4} The $\alpha$-Si$_{3}$N$_{4}$ and $\beta$-Si$_{3}$N$_{4}$ can be obtained using conventional synthesis conditions. Both phases are in the hexagonal crystal system. In 1999, the cubic spinel structure ($\gamma$-Si$_{3}$N$_{4}$) was synthesized by applying high pressure with high temperature.\cite{1999NatureX3N4} The superior physical and chemical properties of wide-gap semiconductors make these materials an ideal choice for device fabrication for many applications, including light emitters, high-temperature and high-power electronics and microwave devices, and micro-electromechanical systems. The effects of temperature on these materials and devices still remain of critical interest.\cite{IEEE1,IEEE2} In addition, wide-gap materials are being considered as alternative gate oxides in the electronic industry, and for low-loss dielectric waveguides in integrated optics,\cite{1999NatureX3N4} which also find its use in light-emitting devices too.\cite{Volodin1997,Steimle2003,Choi2005}

{\par}Here, we present an electronic-structure study of wide band-gap silicon-nitride polymorphs. We use spin-polarized version of the van Leeuwen and Baerends (LB)\cite{VLB} exchange correction to  local density approximation (LDA).\cite{vBH} The LB-corrected LDA is implemented within the framework of tight-binding linear muffin-tin orbital method.\cite{TBLMTO,PS2016}
The LB correction\cite{VLB}  to the exchange potential corrects the asymptotic regions of finite systems, such that the potential behaves as $-1/r$ at large $r$. For an extended system, asymptotic behavior is expected within the interstitial region between atoms.\cite{PS2016} Expectedly, this improves the valence and conduction band energies of an extended system leading to better band gaps.\cite{PS2016,PS2013} The idea is based on the results of calculations  employing the optimized potential method\cite{OPM} and the Harbola-Sahni potential.\cite{HSP} For polymorphs of silicon-nitrides, the LB-corrected LDA shows significantly large improvement over LDA and GGA calcualtions, and found to be similar to hybrid and exact exchange functions.\cite{PS2016} We discuss, how the better exchange-correlation provides improved treatment without explicit need for self-interaction corrections.\cite{SIC1,SIC2,SIC3} 

In Sec.~II, we describe computational methodology. The crystal structure and mechanical properties of Si$_{3}$N$_{4}$ are discussed in Sec.III. The electronic-structure properties of are discussed in Sec.IV.  We summarize our results in Sec.~V, and provide an Appendix with discussion of errors and improvements for three single-electron examples from exact, LDA, and LDA+LB potentials.

\section{COMPUTATIONAL DETAILS}
Density functional theory calculations are performed using the TB-LMTO method within the atomic sphere approximation (ASA).\cite{TBLMTO} In TB-LMTO, core states are treated as atomic-like in a frozen-core approximation and  higher-lying valence states are addressed in the self-consistent-field method applied to the effective crystal potential. This is constructed by overlapping spheres for each atom in the unit cell. It uses two-fold criteria for generating crystal potential: $\left(a\right)$ use of linear combinations of basis functions like plane waves in the nearly free-electron method, and $\left(b\right)$ matching condition for partial waves at the muffin-tin sphere.\cite{TBLMTO}
The open-shell structure of semiconductors and the use of the ASA within the TB-LMTO\cite{TBLMTO} requires us to divide unit cell into atomic spheres (AS) and empty spheres (ES) to fill space for improved basis sets. Here, ES are sites with no cores and  small density of electronic charges.

Next, we utilize the LB-correction to LDA-exchange potential\cite{VLB} to improve the local potential associated with the $-1/r$ behavior of the exchange potential at large distances from the atomic core, which is the interstitial region between neighboring atoms. The LB-corrected LDA provides a dramatic improvement in the band gaps, as well as how it compares, and its relation to, pseudopotential methods that require, in addition, the static dielectric constant.\cite{PS2016, PS2013} Details and further background are available. \cite{HSP,PS2015book,HSH2014} ~The contribution of exchange in empty spheres is very small; thus, any correction to exchange will be even smaller, so we implement the LB-correction in the ASA spheres only. Details of the LDA potentials (correlation V$_{c}$ and exchange V$_{x}$) and the LB-correction to these are given in the Appendix for a single-electron system for potentials with different boundary conditions, and identify and clarify the origin of the improvements to  total energy of the  system,  reducing the error in energy from LB-corrected LDA to less than 10\%, compared to 30\% for LDA, effectively reducing any for self-interaction corrections.

All calculations are done self-consistently (utilizing Anderson mixing\cite{Amix1965} of charge densities) and non-relativistically for given experimental geometry until the ``averaged relative error'' to the previous iteration reaches 10$^{-5}$ for the charge density and 10$^{-4}$ Ryd/atom for the energy. We use LDA+LB corrected exchange in combination with correlation potential parametrized by van Barth-Hedin.\cite{vBH} The k-space integration over occupied electronic states is performed useing the tetrahedron method on a $12\times12\times12$ for cubic cells, $6\times6\times12$, and $6\times6\times18$  set of k-point mesh. In all calculations we chose $R_{ASA}$ to be within 5\% to 10\% of the
default value.\cite{TBLMTO}

\begin{figure*}[t]
\includegraphics[scale=0.4]{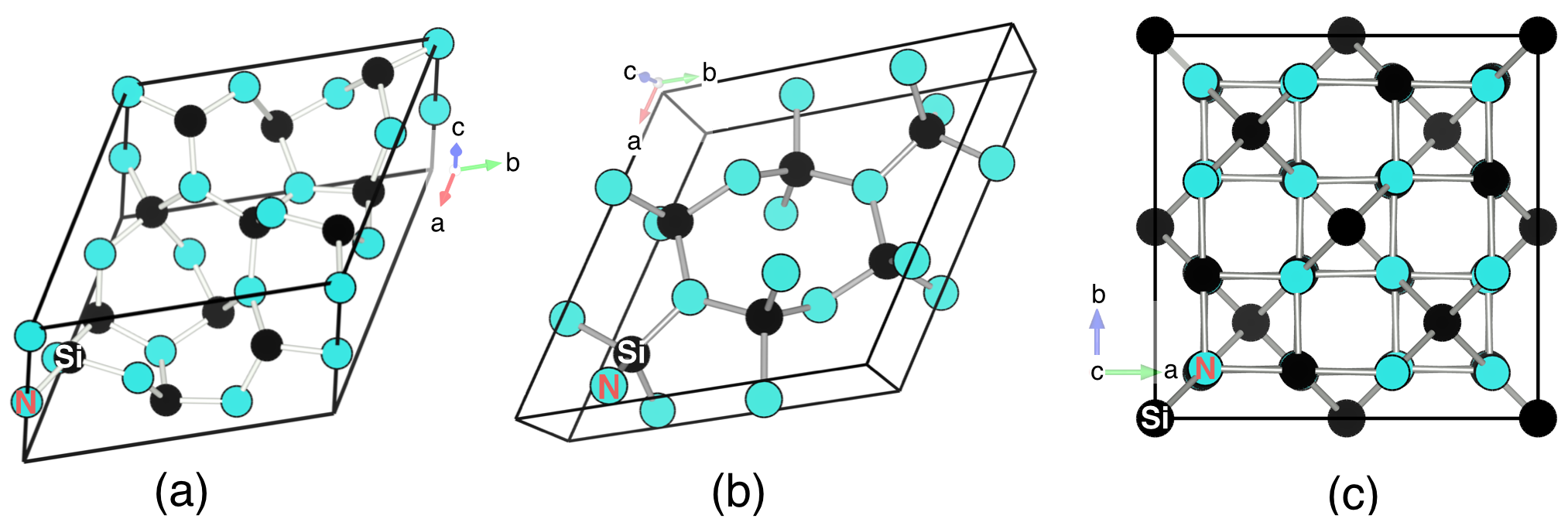}
\caption{(Color online). Crystal structures of (a) trigonal $\alpha$-Si$_{3}$N$_{4}$ (space group: 159), (b) hexagonal $\beta$-Si$_{3}$N$_{4}$ (space group: 173), and (c) cubic-Si$_{3}$N$_{4}$ (space group: 227).}
\label{fig1}
\end{figure*}

\section{Crystal structure} 

{\bf $\alpha$, $\beta$ and $\gamma$-phase Si$_{3}$N$_{4}$:} We show three crystallographic structures of Si$_{3}$N$_{4}$ in Fig.\ref{fig1}, designated as $\alpha$, $\beta$ and $\gamma$-phases.\cite{Leitch2004} The  $\alpha$ and $\beta$ phases are the most common forms, which can be produced under normal pressure condition. The $\gamma$ phase can only be synthesized under high pressures and temperatures and has a hardness of 35 GPa.\cite{Leitch2004,Jiang2001} The crystal structures and unit-cell parameters of $\alpha$, $\beta$, and $\gamma$-Si$_{3}$N$_{4}$  have been extensively studied by many.\cite{Boulay2004,Turkdogan1958,Hardie1957,Ruddlesden1958, Marchand1969,Kohatsu1936,Kato1975,Billy1983,Schneider1994,Yang1995,Toraya2000} 

The $\alpha$- and $\beta$-phase have trigonal (Pearson symbol $hP28$, space group $P31c$, No. 159), and hexagonal ($hP14$, $P6_{3}$, No. 173) structures, respectively, which are built up by corner-sharing SiN$_{4}$ tetrahedra. They can be regarded as consisting of layers of Si and nitrogen atoms in the sequence ABCDABCD$\dots$ or ABAB$\dots$ in $\alpha$- and $\beta$-phase, respectively. The AB layer is the same in the $\alpha$ and $\beta$ phases, and the CD layer in the $\alpha$ phase is related to AB by a c-glide plane. The tetrahedra's in $\beta$-phase are interconnected in such a way that looks like tunnels, running parallel with the c-axis of the unit cell. Due to the c-glide plane that relates AB to CD, the $\alpha$ structure contains cavities instead of tunnels.  An essential difference between the two phases is that the lattice constant in the c$-$axis in $\alpha$-phase is approximately twice that in $\beta$-phase. \cite{RMP1992,Teter1996} 

In Fig.\ref{fig1}(a)\&(b), the Si-atom in $\alpha$ and $\beta$-phase coordinates with four N atoms, forming a SiN$_{4}$ tetrahedron, however, a N-atom is coordinated with three X-atoms, forming a NSi$_{3}$ triangle. These features suggest Si-sp$^{3}$ and N-sp$^{2}$ hybridizations for the nearest silicon and nitrogen pairs.\cite{Yashima2007}

The $\gamma$-phase has spinel structure (cubic, space group $Fd\bar{3}m$, No. 227), which is often designated as c-modification in the literature in analogy with the cubic modification of boron nitride (c-BN).\cite{1999NatureX3N4} It has a spinel-type structure in which two Si-atoms each coordinate six nitrogen atoms octahedrally, and one Si atom coordinates four nitrogen atoms tetrahedrally.\cite{Jiang2001} The cubic-phase comprises of alternating SiN$_{4}$ tetrahedrons and SiN$_{6}$ octahedrons in 1:2 ratio, which are sequentially connected to one another at the N$-$corners in 3D extension. 

The sixfold coordination of Si with N results in a significantly closer atomic packing, associated with a density almost 26\% higher than that of the hexagonal phases.\cite{cSi3N4} The $\alpha$-, $\beta$- and $\gamma$-phase contains 28, 14, and 48 atoms per unit cell, respectively. 

Structural data of $\alpha$-, $\beta$- and $\gamma$-phase Si$_{3}$N$_{4}$ are taken from different high-resolution synchrotron powder diffraction data.\cite{alpha_Cry,beta_Cry,gamma_Cry}

\section{Results and discussion}

\subsection{Structural Properties of Si$_{3}$N$_{4}$-polymorphs}\label{strucProp}

\begin {table}[b]
\caption{Lattice constant and Bulk modulus of Si$_{3}$N$_{4}$ polymorphs obtained from a Birch-Murnaghan equation of state analysis of data points calculated using LB-corrected LDA exchange-correlation potential. We compare LB calculated result with experiments\cite{gammaBM,alphaBMexpt,betaBMexpt} and other theory.\cite{gammaBM,alphaBM,betaBM} The calculated  c/a of  0.73206 ($\alpha$) and 0.36948 ($\beta$) compares well with experiments 0.72497 ($\alpha$) and 0.38262 ($\beta$), respectively.}
\begin {tabular}{cccccccccc}\hline \hline
Phase & \multicolumn{7}{c}{ a(\AA)} {B (GPa)}\\ 
 Si$_3$N$_4$          &  Expt.     &   LB    &   &  LDA   && Expt.  & LB & LDA\\ \hline   
$\alpha$ & 7.752 &  7.819  & &  7.586  &&  248     & 275.6 & 257   \\
$\beta$   & 7.608  &  7.968   & &  7.586  &&  256     &  235.5 & 282     \\
$\gamma$  &  7.742  &  7.681  &&  7.311  &&  300   &  272.8  & 308 \\
\hline \hline
\end {tabular}
\label{tab2}
\end {table}

We calculate structural properties of Si$_{3}$N$_{4}$-polymorphs using LB+LDA exchange-corrected potential. The calculated structural properties are summarized in Table ~\ref{tab2}. The lattice constants of $\alpha-$Si$_3$N$_4$ are determined as a$_{0}$ = 7.819 $\AA$ and c$_{0}$ = 5.724 $\AA$. The $\alpha-$phase is built up from a unit cell with a$_{0}$ = 7.968 $\AA$ and  c$_{0}$ = 2.944 $\AA$, and the lattice constant of the cubic $\gamma-$phase is a$_{0}$  = 7.681 $\AA$. As shown in Table~\ref{tab2}, the lattice constants of all three crystalline silicon nitride phases are in good agreement with experimental results.\cite{gammaBM,alphaBMexpt,betaBMexpt} We obtained the structural properties by fitting the calculated total energies to the Murnaghan\cite{MurnBM} and Birch\cite{BirchBM} equations of state, the calculated bulk modulus is in good agreement with the measured values.\cite{gammaBM,alphaBMexpt,betaBMexpt} This good agreement for the bulk modulus indicates that, to a good approximation, the lattice in Si$_{3}$N$_{4}$-polymorphs undergoes uniform compression with the applied hydrostatic pressure.

\begin{figure*}%[t]
\includegraphics[scale=0.5]{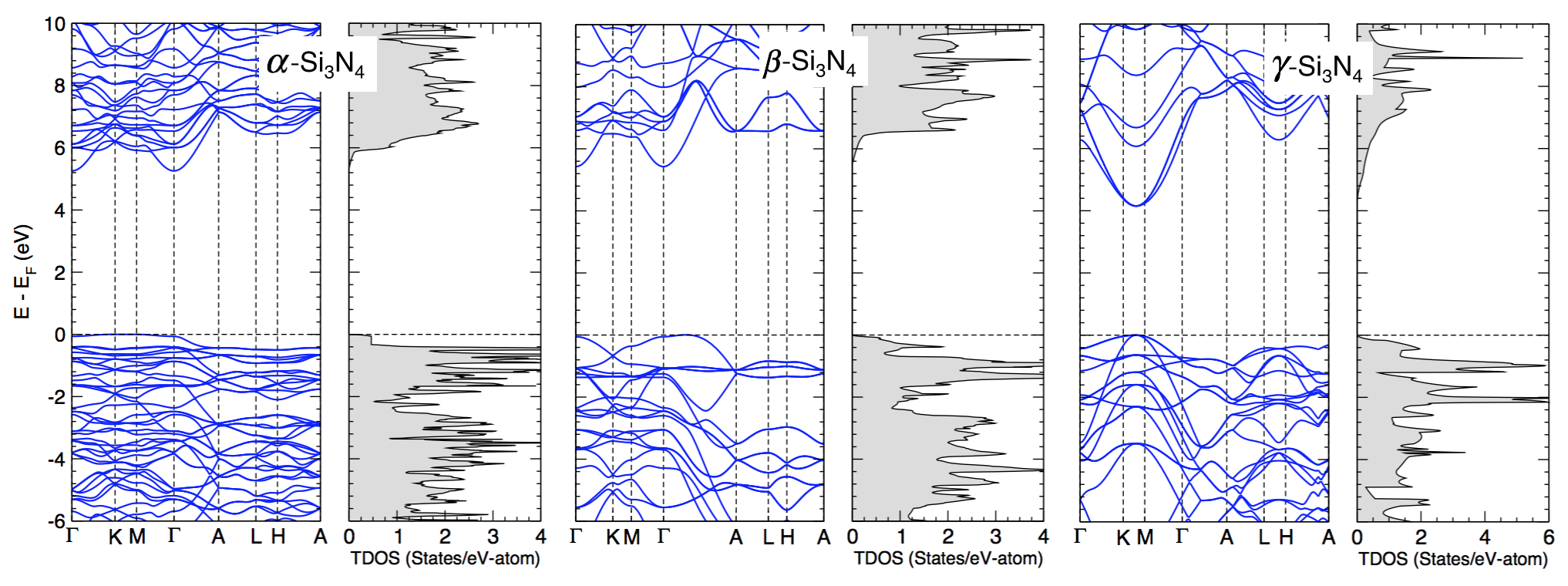}
\caption{(Color online). Electronic structure of silicon-nitride polymorphs, (a) $\alpha-$Si$_{3}$N$_{4}$ , (b) $\beta-$Si$_{3}$N$_{4}$, and (c) $\gamma-$Si$_{3}$N$_{4}$. From the $\Gamma$-point, one can see that only $\gamma$-phase is a direct band-gap material, however, other phases shows indirect band-gap (in agreement with observations).\cite{alphaBG,gammaBG,thesisPBESi3N4} The energies are measured from the top of the valence band.}
\label{fig2}
\end{figure*}

\subsection{Electronic structure of Si$_{3}$N$_{4}$ polymorphs} 

We calculate energetics of Si$_{3}$N$_{4}$-polymorph at lattice constants shown in Table~\ref{tab2}. In terms of energetics, the $\alpha-$ and $\beta-$phases have a very similar energies at the minimum, the energy difference between these two naturally occurring phases is around 30 meV per formula unit. However, the $\gamma-$phase has higher ground state energy than other two phase. Out of the three crystalline silicon nitride phases, the $\beta-$phase has the lowest ground state energy, and unambiguously its thermodynamically most stable phase.\cite{1999NatureX3N4}  

In Fig.\ref{fig2}, we show band-structure and density of states calculated at minimum lattice-constant as described in Sec.~\ref{strucProp}. The calculated band structures are in good agreement with measurements.\cite{gammaBG,Expt1975,Expt1984,Expt1987} The valence band width has two split regions separated by 3$-$4 eV, the lower and upper bandwidths of the valence band are 4$-5$ eV and 10$-$12 eV, respectively. The calculate valence and conduction band-widths are in both quantitative and qualitative agreement with the measured valence and conduction bands from the X$-$ray photoemission\cite{Expt1975,Expt1984,Expt1987}, soft$-$x$-$ray emission \cite{alphaBG}, and bremsstrahlung isochromat-spectroscopy.\cite{Expt1987} The top of the valence band occurs along K$-$M, $\Gamma-$A, and $\Gamma-$A  lines in $\alpha-$, $\beta-$, and $\gamma-$Si$_{3}$N$_{4}$, respectively, while bottom of conduction band is located along K$-$M, at $\Gamma$, and $\Gamma$, respectively. These polymorphs show indirect band gap. Table~\ref{tab1} summarizes the calculated band gaps. For $\alpha$ and $\beta$-Si$_{3}$N$_{4}$, we find reasonably good agreement with the previous calculations.\cite{Kresse} This shows that better treatment of the exchange in the asymptotic region improves the band-gap significantly. For better description of the potential, we provide a discussion of the LB-correction in the appendix for  simple single-electron models, which reduces the error from the exact total energy due to associated self-interaction.

In $\alpha-$Si$_{3}$N$_{4}$, the direct band gap of 5.2 eV at $\Gamma$ is comparable to that  in $\beta-$Si$_{3}$N$_{4}$, which closely follow direct band-gap of $\gamma-$Si$_{3}$N$_{4}$ ($\sim$6.0 eV). Amorphous phase of Si$_{3}$N$_{4}$ are large band-gap materials while in previous calculations cubic phase has been reported to be smaller band gap system than other polymorphs. However in present calculation, we found $\gamma$ phase of Si$_{3}$N$_{4}$ has band-gap of similar order as the other two polymorphs, and in agreement with experiments.\cite{gammaBG}

Comparing total electronic density of states (DOS) of $\gamma$$-$Si$_{3}$N$_{4}$, $\alpha-$Si$_{3}$N$_{4}$, and $\beta-$Si$_{3}$N$_{4}$ shows that $\alpha-$Si$_{3}$N$_{4}$ has a larger band gap, and a steeper VB/CB edge, compared to the other two phases. In $\alpha-$Si$_{3}$N$_{4}$, the bands  around  the  Fermi energy  form a so-called  flat/steep  bands. The flatness  of  the bands  is  very  sensitive  to  the  lattice  patterns and  orbital  interactions in real atomic structures. The polymorphs have very different multiple structures in the DOS in both the VB and CB regions. In Fig.~\ref{fig2}, the electronic structure of silicon-nitride polymorphs split  to form the separate valence band (near -10 eV; not shown in Fig), which reflects the presence of smaller asymmetric part of the potential in Si$_{3}$N$_{4}$. This allows stronger mixing between the low-lying N and C states.

To investigate the bonding nature of $\alpha$ and $\beta$-Si$_{3}$N$_{4}$ in more detail, we calculated and compared the charges around the Si and N sites using the experimental electron-density map. Here, the charges were calculated from the total electronic charges enclosed in atomic-spheres around Si/N-sites. The radii of the spheres were chosen so as to touch both spheres at the density minimum position of the Si$-$N bond.  The calculated atomic charges around the inequivalent Si (Si1, Si2), and N (N1, N2, N3, N4) sites are (10.08, 10.18), and (7.52, 7.99, 7.89, 7.71), respectively, in primitive unit cell of $\alpha$-Si$_{3}$N$_{4}$. Ideally, $\alpha$-phase should have total of 56 e$^{-}$'s, however, we find 51.37 e$^{-}$'s on (Si, N) atomic spheres and 4.63 e$^{-}$'s go into empty-spheres (ESs). Around $\sim8\%$ of charge goes into ESs.  The electrons on Si-AS (N-AS) is lower (higher) than that of the neutral Si-atom (N-atom). These results give direct experimental evidence for charge transfer from the Si to N atoms in Si$_{3}$N$_{4}$-polymorphs. The DFT valence charge density of $\alpha$, $\beta$, and $\gamma$-Si$_{3}$N$_{4}$ have indicated the charge transfer from the silicon atoms to nitrogen atoms. This is mainly ascribed to the difference in the electronegativity. The nitrogen atom is more electronegative than the silicon atoms. In the similar charge analysis for $\beta-$ and $\gamma-$phase, we find $\sim$7\% (AS=19.65 e$^{-}$; ES=1.45 e$^{-}$) and $\sim$6.1\% (AS=32.88 e$^{-}$; ES=2.12 e$^{-}$) of electronic-charge goes on ESs. Keeping in mind 5-10\% overlap of atomic-spheres the amount of charge found in ESs is reasonably good.

\begin {table}[b]
\caption{LB-corrected and LDA band gaps calculated for Si-nitrides polymorphs compared to values from LDA,\cite{alphaBG,gammaBG}, other theories,\cite{Kresse} PBE,\cite{thesisPBESi3N4} and experiments.\cite{Expt1975,Expt1984,Expt1987,alphaBG,gammaBG} }
\begin {tabular}{cccccccccc}\hline \hline
Phase& \multicolumn{7}{c}{Band Gap (eV)}\\ 
Si$_3$N$_4$ &  Expt.     &   LB    &   &  LDA   && Other  & PBE   \\ \hline     
$\alpha$         &  $~$5.0 &  5.24  & &  4.2  &&  ---     &  4.71          \\
$\beta$           &  4.6$-$5.5 &  5.48  & &  4.37  &&  5.92     &  4.45 \\
$\gamma$      &  $~$4.30  &  4.15  &&  3.21  &&  3.45   &  3.38      \\
\hline \hline
\end {tabular}
\label{tab1}
\end {table}

It has been suggested before \cite{Si3N4Bonding,Si3N4Bonding1} that $p-d~\pi$ interaction between Si and N stabilizes the 120$^{\circ}$ Si$-$N$-$Si  in $\alpha-$ and $\beta-$phase Si$_{3}$N$_{4}$ compared to the smaller bond angles found in other trivalent group-V systems. However, we do not find any-significant Si-$d$ orbital contribution in the valence-band states. However, N being a first-row element provides possible explanation for the planar triply-coordinated N-site. The first-row elements, with localized 2$p$ states, behave differently from the elements lying below them in that sp$^{2}$ hybridization is energetically favorable for some first-row solids. The insulator silicon nitride forms crystals with predominantly covalent character and due to the presence of nitrogen, an element from group-III, which leads to hybrid orbitals between Si and N are formed and which differ considerably from the original atomic orbitals.

\section{Conclusion}
We show that the use of LB corrected LDA-exchange for the wide band-gap silicon-nitride polymorphs improves the structural-properties (e.g. lattice constant and bulk-modulus) and electronic-properties  (e.g., the band-gap) significantly. The LB-corrected LDA exchange provides improvement to the exchange-potential over the entire region except near (r$\rightarrow$0).  The LB unerringly corrects the self-interaction in the LDA without need of using SIC-LDA. We show that in existing local-density based exchange-correlation potentials, the self-interaction of correlated electrons is not complete and LB-correction can be used as an  alternative scheme. 

%\section{ACKNOWLEDGMENT}
\emph{Acknowledgements:} The research at Ames Laboratory was supported by the U.S. Department of Energy (DOE),  Office of Science, Basic Energy Sciences, Materials Science and Engineering Division. Ames Laboratory is operated for the U.S. DOE by Iowa State University under Contract No. DE-AC02-07CH11358. 

\vskip 0.5cm

\appendix*

\centerline{\bf APPENDIX} \label{appendix}

\begin{figure}[t]
\includegraphics[scale=0.32]{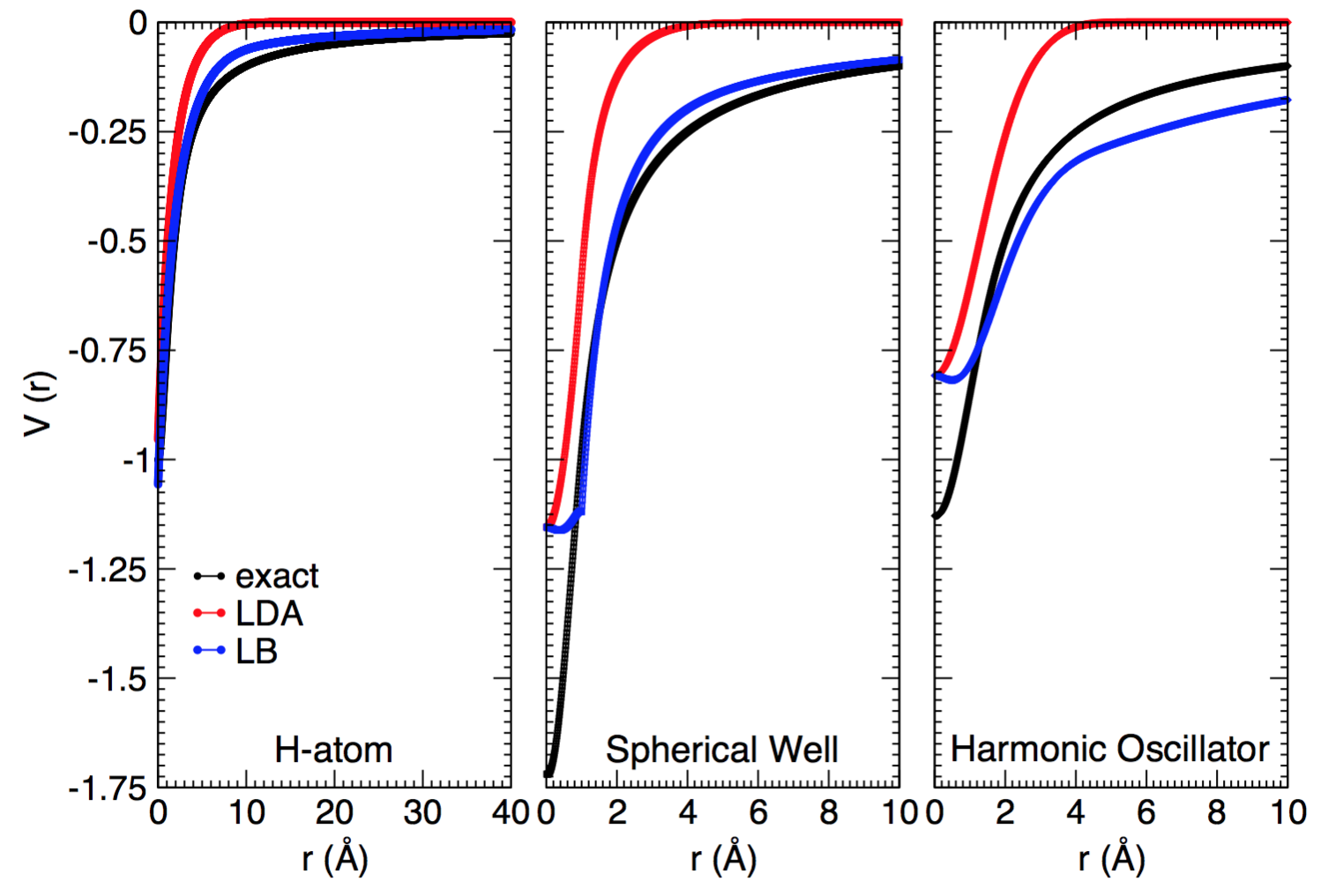}
\caption{(Color online). We plot the exact potential and the LDA potential for a single electron density and compare them with the effective  potential with LB-correction (V$_{x}$+V$_{c}$+V$_{LB}$). Clearly, the LB corrected potential is very close to the exact exchange-correlation potential for both the systems, except very close to $r=0$ limit.}
\label{figapp}
\end{figure}

In the appendix, we describe the LDA potential employed in our calculations along with the LB correction to it.  We show significant reduction in the self-interaction error in the LDA potentials that is brought about by the LB correction.  The spin-polarized exchange and correlation potentials in the LDA are given in terms of the spin density $\rho_{\sigma}$ by the expressions
\begin{equation}
V_{x} = -\left(\frac{6\rho_{\sigma}}{\pi}\right)^{1/3}
\end{equation}
and
\begin{equation}
V_{c} = -0.0254\times\ln\left(1+\frac{75.59}{r_{s}}\right)
\end{equation}
where r$_{s}$ is the Wigner-Seitz radius corresponding to $\rho_{\sigma}$, $\sigma$ is spin-component.  The LB-corrected potential is given in terms of the density and its gradient as
\begin{equation}
V_{x,\sigma}^{LB} ({\bf r}) = -\beta\rho_{\sigma}^{1/3}\frac{x_{\sigma}^{2}}{1+3{\beta x_{\sigma}}{\sinh}^{-1}(x_{\sigma})},
\label{VLB}
\end{equation}
where $\beta=0.05$ was used in the original  formulation.\cite{VLB} The variable $x_{\sigma}=|\nabla\rho_{\sigma}({\bf r})|/{\rho_{\sigma}^{4/3}({\bf r})}$ signifies the change in mean electronic distance provided density is a slowly varying function in given region with strong dependence on gradient of local radius of the atomic sphere $R_{ASA}$. 

\begin{table}[b]
\begin{tabular}{|c|c|c|c|c|c|c|c|ccccc}
\hline
  		System	 &	$V_{exact}$  & $V_{LB}$  &$\frac{(exact-LB)}{exact}$  &	$V_{LDA}$ & $\frac{(exact-LDA)}{exact}$\tabularnewline
		  			 &	 &   &  \% change &	& \% change \tabularnewline
\hline
\hline
H Atom 	&	0.3144 &	0.2858 & 9.09  & 0.2226 & 29.20  \tabularnewline
    Well	& 	0.6240 &	0.5513 &11.65 & 0.4052 & 35.06  \tabularnewline
Oscillator 	&	0.4021 &	0.3720 &  7.49 &0.2750 &	31.61  \tabularnewline
\hline 
\end{tabular}
\caption{ {\label{tab3}}   Total energy (from \ref{A4}) for a single electron in a hydrogen atom, spherical well, and iharmonic oscillator obtained using the exact, LDA and  LB-corrected LDA potentials. The LB-corrected LDA potential provides a significant improvement to the energy without invoking need for self-interaction corrections.} 
\end{table}

The LB potential goes as $-1/r$ for densities that decay exponentially as a function of r. The effective potential V$_{x}$+V$_{c}$+V$_{LB}$ along with the LDA potential V$_{x}$+V$_{c}$ are shown in Fig.~\ref{figapp} for an electron in a hydrogen atom, in a spherical well of depth 5 a.u. and radius 1 a.u. and in a harmonic potential with $\omega$ = 1.0 a.u.  We note that while the density of the electron in the first two systems decays as exp(-${\bf r}$) for positions far from the origin, in a harmonic oscillator the density decays as exp(-${\bf r}^2$). The exact known densities for these systems are used to obtain these potentials. The potentials are compared with the exact exchange-correlation potential given by negative of the electrostatic potential for these single-electron systems.  It is clear that except close to $r=0$, the LB-corrected potential is very close to the exact exchange-correlation potential for both the systems. 

The proximity of the LB-corrected potential to the exact one also leads to much smaller error in self-interaction energy. In the Table~\ref{tab3}, we have displayed the exact potential (self-interaction free) for a single electron systems given by 
\begin{equation}\label{A4}
E_{exact} = \frac{1}{2} \int{d{\bf r}}V_{exact}({\bf r })\rho({\bf r})
\end{equation}
 and compare exact energies calculated with V$_{exact}$ potential with approximate exchange-correlation functional i.e., V$_{LDA}$ or V$_{LB}$, for the same potentials shown in Fig.~\ref{figapp}.  To calculate the self-interaction energy in the approximate potentials, we have used the expression in \ref{A4} instead of the exchange-correlation energy functional.  This is because for single electron systems the exact exchange-correlation energy is indeed given by \ref{A4}.

In Table~\ref{tab3}, the error due to the self-interaction in the LDA is about 30\% in all three systems, compared to the LB-corrected results where it is reduced by a factor of 3 (about 10\%). The LB-corrected LDA shows significant improvement in total-energy without the need of separate treatment of self-interaction.\cite{SIC1,SIC2,SIC3}

\end{document}